\documentclass[10pt,letterpaper]{article}
\usepackage[OE]{express}

\usepackage{siunitx}
\DeclareSIUnit\decibelc{dBc}
\usepackage{amsmath,subfigure}
\allowdisplaybreaks
\usepackage{multirow}

\begin{document}
\title{Self-pulsing in single section ring lasers based on Quantum Dot materials: theory and simulations}
\author{Lorenzo Luigi Columbo,$^{1,2*}$, Paolo Bardella,$^1$ and Mariangela Gioannini$^{1}$}





\address{
$^1$ Dipartimento di Elettronica e Telecomunicazioni, Politecnico di Torino, Corso Duca degli Abruzzi 24, Torino, IT-10129, Italy\\
$^2$ Consiglio Nazionale delle Ricerche, CNR-IFN, via Amendola 173, Bari, IT-70126, Italy}
\begin{center}
\email{*lorenzo.columbo@polito.it} 
\end{center}

\begin{abstract}

We studied theoretically coherent phenomena in the multimode dynamics of single section semiconductor ring lasers with Quantum Dots (QDs) active region.
In the unidirectional ring configuration our simulations show the occurrence of self-mode-locking in the system leading to ultra-short pulses (sub-picoseconds) with a THz repetition rate. As confirmed by the Linear Stability Analysis (LSA) of the Traveling Wave (TW) Solutions this phenomenon is triggered by the analogous of the Risken-Nummedal-Graham-Haken (RNGH) instability affecting the multimode dynamics of two-level lasers. 
\end{abstract}


\bibliographystyle{osajnl}

\section{Introduction}

Semiconductor lasers operating in self-pulsing (SP) regime are valid alternatives to passive and active mode-locked devices for the generation of ultra-short, high-repetition rate optical pulses for applications to optical information encoding and time resolved measurements of e.g fast molecular dynamics \cite{Sato,Rev01,Liu}. In the frequency domain the SP regime corresponds to an Optical Frequency Comb (OFC), i.e a light emission characterised by equally spaced optical lines with low phase noise and low mode partition noise \cite{Hansch, faistreview}. 
These quite simple self-locked sources have attracted an impressive interest for a number of applications in astronomy, spettroscopy, waveform generation, optical clocks and in the rapidly growing field of high-capacity DWDM optical interconnection where the OFC laser diode feeds the silicon photonics optical modulators to realize a compact and low cost transmitter \cite{Kippenberg555,Delfyett, Chen, Eiselt, link2017, faistreview}.  \\

Respect to conventional Quantum Well (QW) based semiconductor lasers, active devices based on low dimensional materials as QDs and Quantum Dashes (QDashes) draw interest because of their broad spectral gain, small nonlinear dispersion, low threshold current, fast gain recovery time and small integration spatial scale.
Experimental evidences of SP in single section Fabry-Perot (FP) lasers based on QDashes \cite{Gosset} and QDs \cite{Lu09} active materials have been reported. In the case of FP configuration we have recently demonstrated \cite{Bardella2017} that the carrier grating induced by the standing wave pattern (not washed out by diffusion in the QDs case) can explain the broad multi-wavelength optical spectra typically observed in QDs lasers, whereas FWM allows the self-locking of the modes when the laser output power it is high enough. When SP occurs, the pulse repetition rate depends on the FP longitudinal cavity mode separation and for typical FP laser length it stays in the tens of GHz range.\\

The question remains on what happens if the standing wave pattern is not present, as for example in a ring laser configuration where only the clockwise (or counter clockwise) mode propagates. In this case, as shown by some recents works on Quantum Cascade Lasers (QCLs) that share with QDs laser similar dynamical features \cite{FaistBook,Boiko, Gordon}, multi-wavelength emission and self-pulsation should be triggered by a RNGH instability of the TW that consists in the parametric amplification of the cavity modes resonant with the frequency of the Rabi oscillations (Rabi frequency, $\nu_{R}$) of the system \cite{Lugiato, Risken}. The RNGH instability can be considered the epitome of the self-mode-locking in a two-level laser. As happen for unipolar laser such as QCLs where lasing action involves intersubband transitions, in low dimensional active material such QDs and QDashes intraband lasing transitions between discrete levels are associated with quite narrow and symmetric gain linewidth that makes this class of emitters behave quite similarly to two-level lasers and thus it allows for the observation of coherent effects like Rabi oscillations \cite{caNat,Kolarczik, Capua}. On the contrary, in more conventional bipolar semiconductor lasers based on QW the Rabi oscillations and the consequently RNGH are usually hindered due to the quite broad and asymmetric resonance and, to the best of our knowledge, not a single observation of these phenomena have been reported so far .\\
We note that ring lasers and passive resonators are nowadays key elements for the realization of photonic integrated circuits \cite {NatLiu} and unidirectional propagation can be easily obtained \cite{Longhi17, Sorel}. The details of fabrication and characterisation of QDs ring lasers emitting at $1.5$ $\mu m$ are reported for example in \cite{Proc2006}.\\

We propose here an unidirectional QDs ring laser of few millimeters length where multimode emission leads to SP as a result of a RNGH instability of the TW solutions. We also show that this phenomenon in unidirectional ring QDs lasers is reliable on a wide range of bias currents and device lengths. As demonstrated in Section 3 as a consequence of the RNGH instability the SP repetition rate is in the hundred of GHz or few THz range, even if the ring cavity FSR is tens of GHz as in the standard FP laser configuration. \\

In order to simulate the multimode dynamics of the QD ring laser by properly taking into account coherent radiation-matter interaction we extended the Time Domain Travelling Wave (TDTW) model described in \cite{Rossetti, Gioannini} to include the temporal evolution of the medium polarization. We calculated the TW solutions of the system and we studied their stability against spatio-temporal perturbations with a standard Linear Stability Analysis (LSA) technique. As expected, the results of LSA show that the TW instability is associated with the amplification of the Rabi frequency in the QD active material that behaves as ensemble of artificial two-level atoms, thus having a RNGH character. Our numerical simulations show that the system spontaneously evolves towards a multimode solutions that corresponds to ultra-short pulses (hundreds of \SI{}{\femto\second}) at \SI{}{\tera\hertz} repetition rate, close to $\nu_{R}$.\\
The numerics also reveal that an increase of the inhomogenous broadening, that may represents an additional incoherent effect in the multimode competition \cite{Lugiato}, reduces the intervals of the bias current where SP is observed.\\
We finally observe that SP and the consequent OFC with THz or sub-THz optical line spacing can be desirable for a number of applications among which we mention the photonic generation of THz or sub-THz signals by illuminating a fast photodetector with the SP optical signal. This simple ring source could be therefore a valid alternative to the mixing of comb lines used nowadays to generate the THz signal \cite{Koenig, Rev04THz, Rev05THz}. \\

The paper is organised as follows: in Section 2, we describe the TDTW model used for simulating the multimode dynamics of the ring cavity single section QD laser. In Section 3, we present and discuss the results of LSA of the TW solutions and the numerical simulations for standard unidirectional InAs/GaAs QD laser. Finally we draw our conclusions in Section 4.

\section{Multi-populations Time Domain Travelling Wave Model}

\begin{figure}[tb]
\centering
    \subfigure[]{\includegraphics[width=0.4\textwidth]{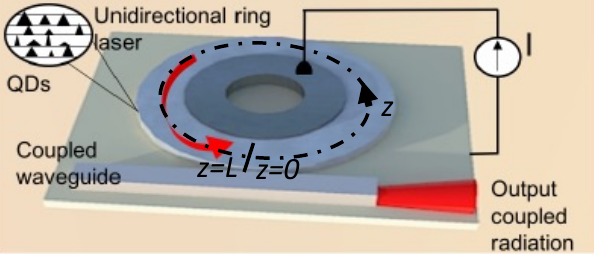}}
      \subfigure[]{\includegraphics[width=0.8\textwidth]{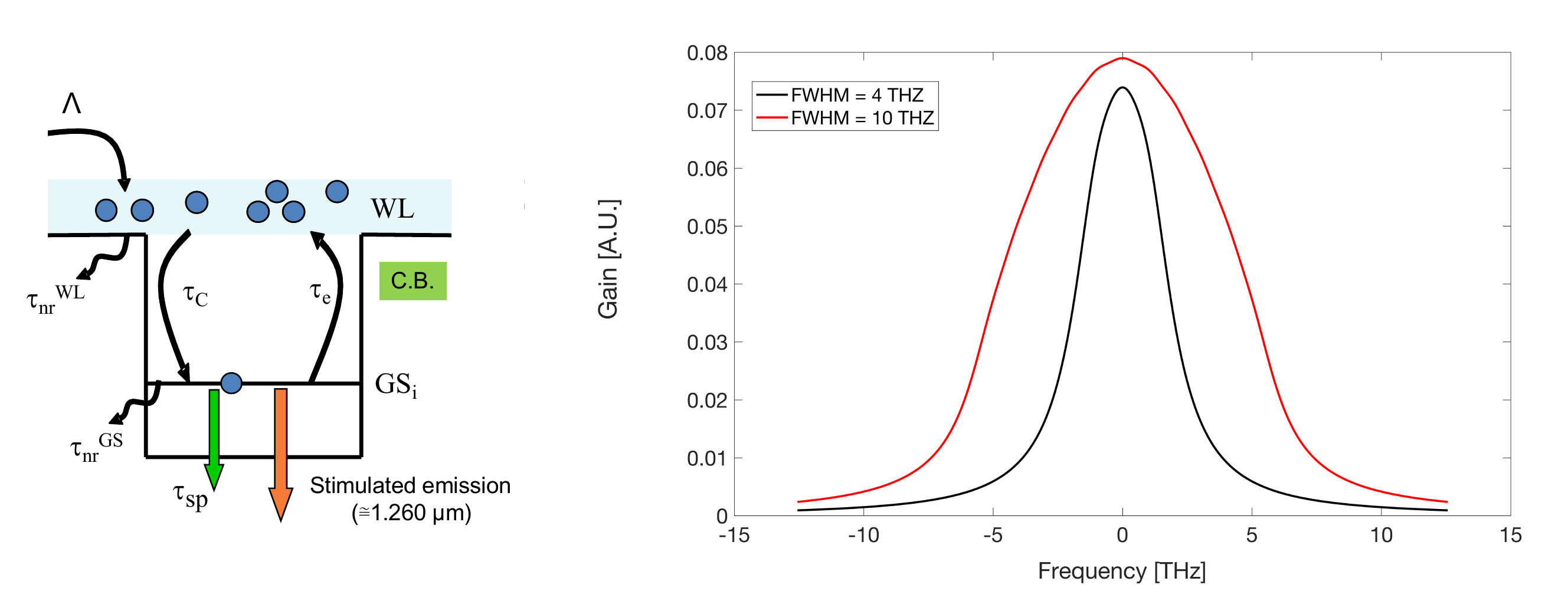}}
    \caption{(a) Sketch of the unidirectional ring configuration. (b) Schematic of the electron dynamics in an exemplary quantum dot sub-group $i$ (left). Effective gain lineshape corresponding to inhomogeneous gain broadening of $\simeq$ \SI{4} {\tera \hertz} ($\simeq$ \SI{16} {\milli \electronvolt}) and $\simeq$ \SI{10} {\tera \hertz} ($\simeq$ \SI{41} {\milli \electronvolt}). The the FWHM of the homogeneous gain linewidth is $2\Gamma$ $\simeq$ \SI{2.4} {\tera \hertz} ($\simeq$ \SI{10} {\milli \electronvolt}) corresponding to a dipole dephasing time of \SI{130}{\femto\second} (right). The zero frequency in the $x$-axis corresponds to $\omega_{0}/(2\pi)$.} 
 \label{fig:3}
\end{figure}

We consider a single section Quantum-Dots-in-a-Well (DWELL) InAs/GaAs ring laser emitting from the ground state (GS) around \SI{1258}{\nano\meter} \cite{Gioannini}. The length of the laser cavity ($L$) is a few hundreds of microns. The laser structure with the coordinate system is sketched in Fig.\ref{fig:3}.a, whereas the QDs states, electron dynamics and gain line shapes for different inhomogeneous broadening are shown in Fig. \ref{fig:3}.b.\\

We sketch in Fig. \ref{fig:3} the electron dynamics as taken in our model. The main material and device parameters are summarised in Table \ref{Tableparam}. 
The coherent interaction between QDs inhomogeneous broadened gain medium and the intracavity electric field is described trough a set of coupled traveling wave equations for the slowly varying envelops of the fundamental TE electric field $E(z,t)$ and of the slowly varying envelop of the microscopic polarizations $p_{i}(z,t)$, coupled with the evolution equations for the electron occupation probabilities of ground state $\rho_{i}$ in each dot group and in the wetting layer (WL) $\rho_{WL}$ \cite{Gioannini}.
\begin{eqnarray}
\frac{\partial E (z,t)}{\partial t} &=&\gamma_{p} \left(-\frac{\partial E}{\partial z}-\frac{\alpha_{wg}L}{2}E-C\sum_{i=-N}^{N} \bar{G}_{i} p_{i} \right)\label{fieldfastc}\\
\frac{\partial p_{i}(z,t)}{\partial t}&=&\left [(j\delta_{i}/ \Gamma-1)p_{i}-D(2\rho_{i}-1)E\right ]    \label{Pfast1b}\\
\frac{\partial \rho_{i}(z,t)}{\partial t}&=& -\rho_{i}\gamma_{e}(1-\rho_{WL}) +F \rho_{WL} \gamma_{C}(1-\rho_{i}) \\
&-&\gamma_{sp}\rho_{i}^{2} -\gamma_{nr}^{GS}\rho_{i}
+ H \, Re\left(E^{*}p_{i} \right)          \label{pop1fastb}\\
\frac{\partial \rho_{WL}(z,t)}{\partial t}&=& \Lambda \tau_{d}-\gamma_{nr}^{WL}\rho_{WL} +  \sum_{i=-N}^{N}\left[-\bar{G}_{i} \rho_{WL}\gamma_{C}(1-\rho_{i}) 
+  \frac{\bar{G}_{i}}{F}  \rho_{i}\gamma_{e}(1-\rho_{WL}) \right] \label{pop2fastb}
\end{eqnarray}
In the convenient adimensional formulation provided by Eqs. (\ref{fieldfastc})-(\ref{pop2fastb}) we scaled time to the fastest time scale in the system represented by the dipole dephasing time $\tau_{d}$ and the longitudinal coordinate to the cavity length $L$. The complex dynamical variables are linked to the corresponding physical quantities by the relations:
$$E\longrightarrow E \sqrt{\frac{\eta}{\Gamma_{xy}}} \frac{d_{GS}}{\hbar \Gamma}, \quad p_{0,i\, GS} \longrightarrow j\, p_{0, i\, GS} 
\sqrt{\frac{\eta}{\Gamma_{xy}}} \frac{N_{D}|d_{GS}|^{2}}{\epsilon_{0}\hbar \Gamma h_{QD}}$$
where $d_{GS}$ is the dipole matrix element associated to the optical transition from ground level, $\eta$ is the effective refractive index, $\Gamma_{xy}$ is the transverse optical confinement factor in the total QD active region, $\Gamma=1/\tau_{d}$, $h_{QD}$ is the QDs layer thickness, $N_{D}$ is the number of QDs per unit area. The adimensional parameters $C$, $D$, $F$, $H$ have the following expressions:
\begin{equation*} 
C= \frac{\omega_{0}L\Gamma_{xy}\mu}{2c\eta}, \,   D=\frac{|d_{GS}|^{2}N_{D}}{\epsilon_{0}\hbar \Gamma h_{QD}}, \, F=\frac{D_{WL}}{\mu N_{D}}, \, \\ 
H=\frac{\tau_{sp}\Gamma^{2} \omega_{0} \Gamma_{xy} \hbar \epsilon_{0} h_{QD}}{\eta \omega_{i \, GS} |d_{GS}|^2 N_{D}}
\end{equation*} 
where $\omega_{0}$ is our reference angular frequency coincident with the cold cavity mode closest to the GS gain peak, $\mu$ is the degeneracy of the ground state, $\omega_{i \, GS}$ is the transition frequency of the $i$ group so that $\delta_{i}=\omega_{i \,GS}-\omega_{0}$, $D_{WL}$ is the number of WL level per unit area per QDs layer and $\tau_{sp}$ is the spontaneous electrons decay time from the GS state. Moreover in the previous equations $\alpha_{wg}$ represents the wave guide losses, $\gamma_{p}=\tau_{d}v_{g}/L$ is the normalized photon decay rate, $\gamma_{e,C}=\tau_{d}/\tau_{e,C}$ are the normalized escape and capture rates, $\gamma_{sp}=\tau_{d}/\tau_{sp}$, $\gamma_{nr}^{WL,GS}=\tau_{d}/\tau_{nr}^{WL,GS}$ represent the normalized nonradiative decay rates and $\lambda$ is the carriers injection probability per unit time. Finally $\bar{G}_{i}$ is the probability that a QD belong to the sub-group $i$ and it follows a Gaussian distribution.\\
\begin{table}
\centering
\begin{minipage}[c]{0.5\textwidth}
\centering
\footnotesize
\setlength\tabcolsep{2pt} 
			\begin{tabular}{ccc} \hline \hline Symbol & Description & Values\\ \hline
				\multicolumn{3}{c}{Material parameters}\\ \hline
				$\eta$&Effective refractive
				index &\SI{3.34}{}\\ 
				
				$\mu$&Confined states degeneracy &\SIlist{2}{}\\
				$1/\Gamma$&Dipole dephasing time&\SI{130}{\femto\second}\\
				$d_{GS}$&Dipole matrix element for GS&\SI{0.6}{\electronvolt \nano\meter}\\
				$\tau_{C}$& Electron capture times &\SIlist[parse-numbers=false]{1}{\pico\second}\\
				$\tau_{e}$& Electron escape times &\SIlist[parse-numbers=false]{1.5}{\pico\second}\\
				$\tau_{nr}^{WL}$& Electron non-radiative decay times &\SIlist[parse-numbers=false]{1}{\nano\second}\\
				$\tau_{nr}^{GS}$& Electron non-radiative decay times &\SIlist[parse-numbers=false]{few}{\nano\second}\\
				$\tau_{sp}$& Electron spontaneous emission time&\SI{2}{\nano\second}\\ \hline 
				\multicolumn{3}{c}{Device parameters}\\ \hline 
				$w$&Ridge width&\SI{5}{\micro\metre}\\ 
				$n_L$&Number of QD layers&\SI{15}{}\\
				$N_D$&QD surface density&\SI{2.7e10}{\per\centi\metre\squared}\\
				$D_{WL}$&Wetting layer electron levels surface density&\SI{2.1e11}{\per\centi\metre\squared}\\		
				$h_{QD}$&QD layer thickness&\SI{5}{\nano\metre} \\ 
				$\alpha_{wg}$&Intrinsic  waveguide losses&\SI{4}{\per\centi\meter}\\ 
				$r$ & Coupler reflectivity&\SI{1}{}\\
				$L$&Device  length&\SI{200}{\micro\metre}\\ $\Gamma_{xy}$& Transverse optical confinement factor&\SI{12}{\percent}\\  \hline
			\end{tabular}	
			\label{Tableparam}	
		\end{minipage}
		\caption{Main materials and device parameters used in the TDTW model.}
			\label{Tableparam}
    \end{table}
%
In the unidirectional ring configuration the field envelope satisfies the boundary condition:
\begin{equation*}
E(0,t)=\sqrt{1-k^{2}}E(L,t), 
\end{equation*}
where $k$ is the output coupling coefficient between the ring and the coupled waveguide (see Fig. \ref{fig:3}.a).
Considering our normalization and the physical constants, the output power, expressed in \SI{} {\milli \watt}, can be obtained by multiplying $|E(z,t)|^{2}$ by a factor of about $35$.
Finally, for sake of simplicity, we consider in this work the case where emission only occurs from the ground state \cite{Gioannini}.

%


\section{Risken-Nummedal-Graham-Haken instability}
To study the character of the TW instability we performed as reported in this section a semi-analytical linear stability analysis. This analysis requires first the calculation of the TW solutions, i.e. the single frequency solutions of Eqs. (\ref{fieldfastc})-(\ref{pop2fastb}) (as detailed in paragraph 3.1) and then the evaluation of the stability against spatio-temporal perturbations of this solutions by calculating the perturbations parametric gain (as detailed in paragraph 3.2). 
\subsection{TW solutions}
We looked for the single frequency solution of Eqs. (\ref{fieldfastc})-(\ref{pop2fastb}) detuned in general of a quantity $\delta \omega$ from the gain peak $\omega_0$ in the form $$E=\overline{E}e^{j (\delta \omega/\Gamma \,t - \delta k \,L z)}, \quad  p_{i}=  \overline{p_{i}}e^{j(\delta \omega/\Gamma \, t- \delta k \, L z)}$$ $$\quad \rho_{i}=  \overline{\rho_{i}},\quad  \rho_{WL}=  \overline{\rho_{WL}}$$ where we set $\delta k = \delta \omega /v_{g}$ and we introduced the group velocity $v_{g}=c/\eta$. This led to: 
\begin{eqnarray}
\overline{p_{i}}&=&\frac{ \left[D(2\overline{\rho_{i}}-1)\overline{E}\right] }{j\delta_{i}/ \Gamma-1-j\delta \omega /\Gamma } \label{Pfastaz}\\
\overline{\rho_{WL}}&=&\frac{\Lambda \tau_{d}+\frac{1}{F}\sum_{i=-N}^{N}\bar{G}_{i}  \overline{\rho_{i}}\gamma_{e}}{\gamma_{nr}^{WL}+\sum_{i=-N}^{N}\bar{G}_{i}\gamma_{C}(1-\overline{\rho_{i}}) + \frac{1}{F} \sum_{i=-N}^{N} \bar{G}_{i}\overline{\rho_{i}}\gamma_{e}}\label{pop2fbstaz}\\
0 &=& \overline{E} \left(\frac{\alpha_{wg}L}{2}+C \, D \sum_{i=-N}^{N} \bar{G}_{i} \frac{(2\overline{\rho_{i}},-1)}{j\delta_{i}/ \Gamma-1- j\delta \omega /\Gamma}\right)\label{fieldstaz11}\\
0&=&-\overline{\rho_{i}}\gamma_{e}(1- \overline{\rho_{WL}})+F\overline{\rho_{WL}} \gamma_{C}(1-\overline{\rho_{i}}) - \gamma_{sp} \overline{\rho_{i}}^{2} + H \, D Re\left( \frac{|\overline{E}|^2(2\overline{\rho_{i}}-1)}{j \delta_{i}/\Gamma -1-j\delta \omega /\Gamma}  \right)    \label{pop1fstaz}
\end{eqnarray}
While in presence of in presence of inhomogeneously broadened gain and therefore multiple populations the TW solution can be found only by numerically solving the implicit nonlinear equations (\ref{Pfastaz})-(\ref{pop1fstaz}), in case of perfect homogeneous medium (i.e. only one population considered, $i=1$) the TW equations (\ref{Pfastaz})-(\ref{pop1fstaz}) have an analytical solution.

\subsection{Linear Stability Analysis of the TW solutions}

\begin{figure}[ht]
	\centering
	\includegraphics[width=8.cm]{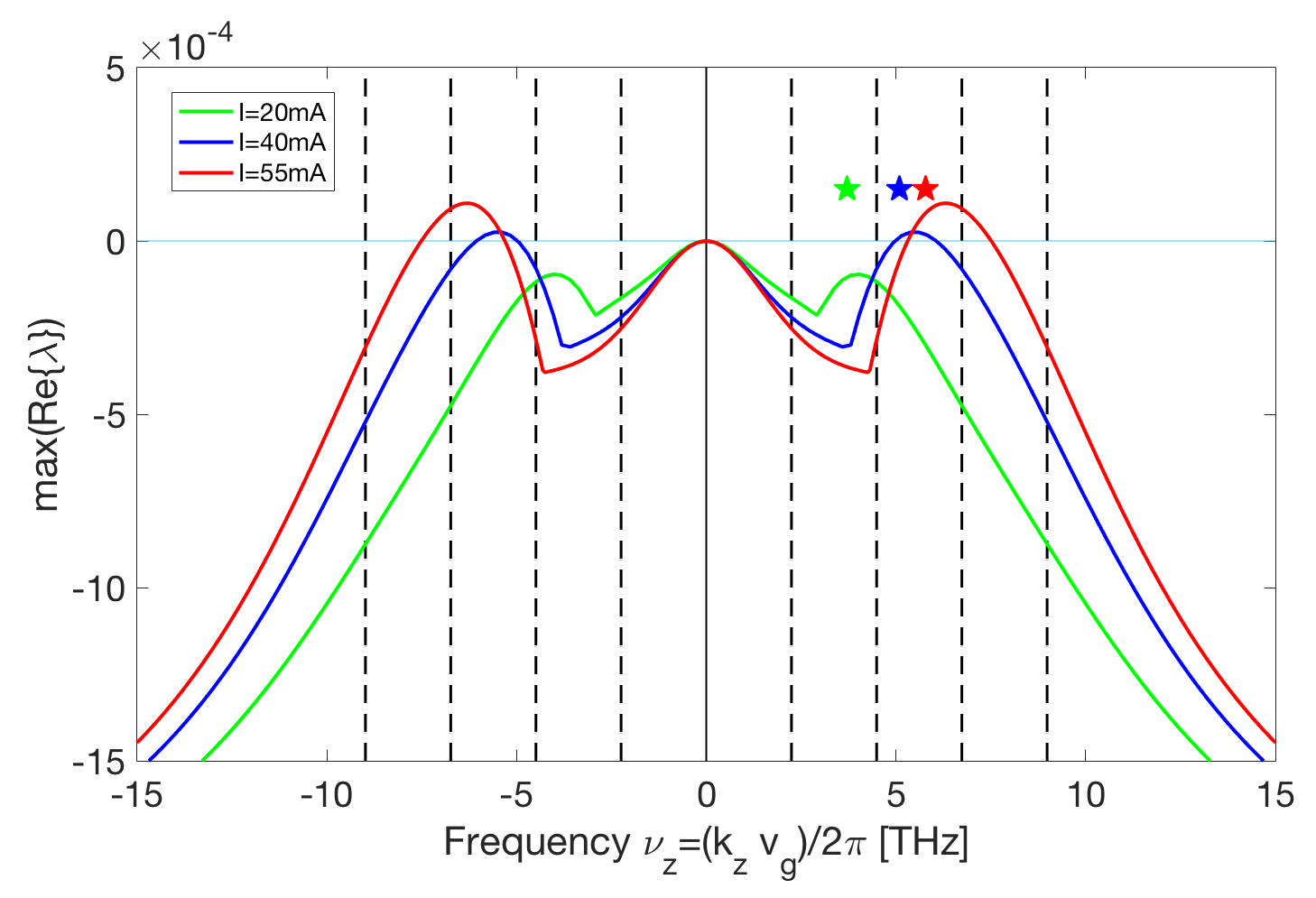}
	\hspace{0mm}\\
		\caption{Results of the LSA of the TW solutions for different bias currents. Plot of the parametric gain for each value of the frequency $\nu_{z}=\omega_{z}/2 \pi=k_{z} \, v_{g}/2 \pi$ treated as continuous variable. Dashed lines indicate the first values of $k_{z}$ compatible with the periodic boundary conditions. We consider $3$ QDs populations resonant with the lasing light associated with central angular frequencies $0$, $1.0$$THz$ and $-1.0$$THz$ that lead to a FWHM of the effective inhomogeneous broaden gain linewidth of $\simeq$ \SI{4} {\tera \hertz} ($\simeq$ \SI{16} {\milli \electronvolt}) (see Fig. \ref{fig:3}). The other parameters are those used in \cite{Gioannini}. The symbols indicate the position of the Rabi frequencies estimated using Eq. (\ref{rabi1eq}) closest to the parametric gain peak.}
        \label{fig0}  
\end{figure}

The LSA of Eqs. (\ref{fieldfastc})-(\ref{pop2fastb}) around the TW solutions is carried out in detail in Appendix A. The parametric gain, i.e. the maximum of the real part of the perturbation eigenvalue $\lambda$ at a frequency $\nu_{z}=\omega_{z}/2 \pi=k_{z}\,v_{g}/2 \pi$ relative to the TW frequency treated as continuous variable, is plotted for e.g. in Fig. \ref{fig0}. The cold cavity modes are those indicated by the dashed lines. We observe that the TW is unstable for $I\ge 55mA$ where a positive parametric gain favour the exponential growing of the modes with $k_{\mp 3}=\mp 3 \times 2 \pi \, /L$. For lower currents in fact the parametric gain for all the cold cavity modes is negative.

\begin{figure}[ht]
	\centering
         \includegraphics[width=0.7\textwidth]{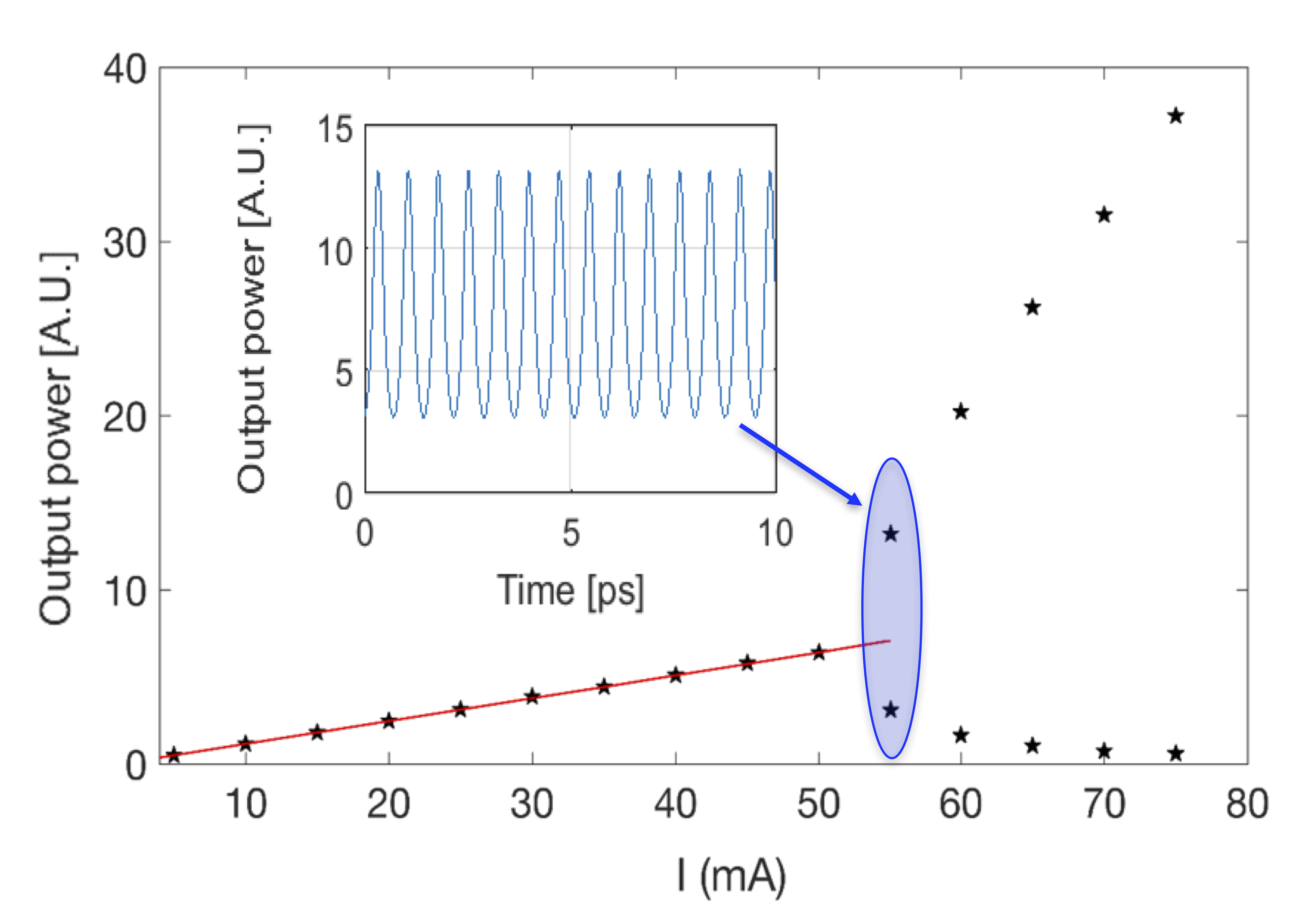}
	\caption{Bifurcation diagram of the TW solutions: the maxima and minima in the intensity time traces are reported against the bias current as control parameter. Red lines correspond to the TW solutions calculated using Eqs. (\ref{Pfastaz})-(\ref{pop1fstaz}).}
        \label{fig1}  
\end{figure}
As shown in Fig. \ref{fig0} we verified that the instability starts by developing Rabi sidebands around the TW lasing frequency and it can be seen as an amplification of the Rabi oscillations, in this sense it can be interpreted as a RNGH instability affecting the TW solutions in multimode two-level atoms \cite{Lugiato}. 

The calculation of the Rabi frequency $\nu_{R}=\omega_{R}/2 \pi$, associated to periodic exchange of energy between light and matter of the system, is reported in the following paragraph and is based on the very well justified hypothesis that the QDs active is to analogous to an ensemble of artificial two-level atoms.\\
Using the standard method described for example in \cite{Lugiato} we calculate the Rabi frequencies associated with the each group of QDs of the multi-population ensemble:
\begin{equation}
\nu_{R,i}=\left(\gamma_{sp} H \, D \, 2 |E|^{2}+((\delta_{i}-\delta \omega)/\Gamma)^2\right)^{0.5}/2 \pi
\label{rabi1eq}
\end{equation}




The large value of the $\nu_{R}$, that turns out to be of the order of the inverse of the coherence time ($1/\Gamma$) may also explain the recent experimental observations of Rabi oscillations effect in intense pulse propagation in QDs based SOA at room temperature \cite{Kolarczik, Capua}.

 \section{Results of dynamical simulations}
  \begin{figure}[h!tb]
 \begin{center}
\includegraphics[width=0.8\textwidth]{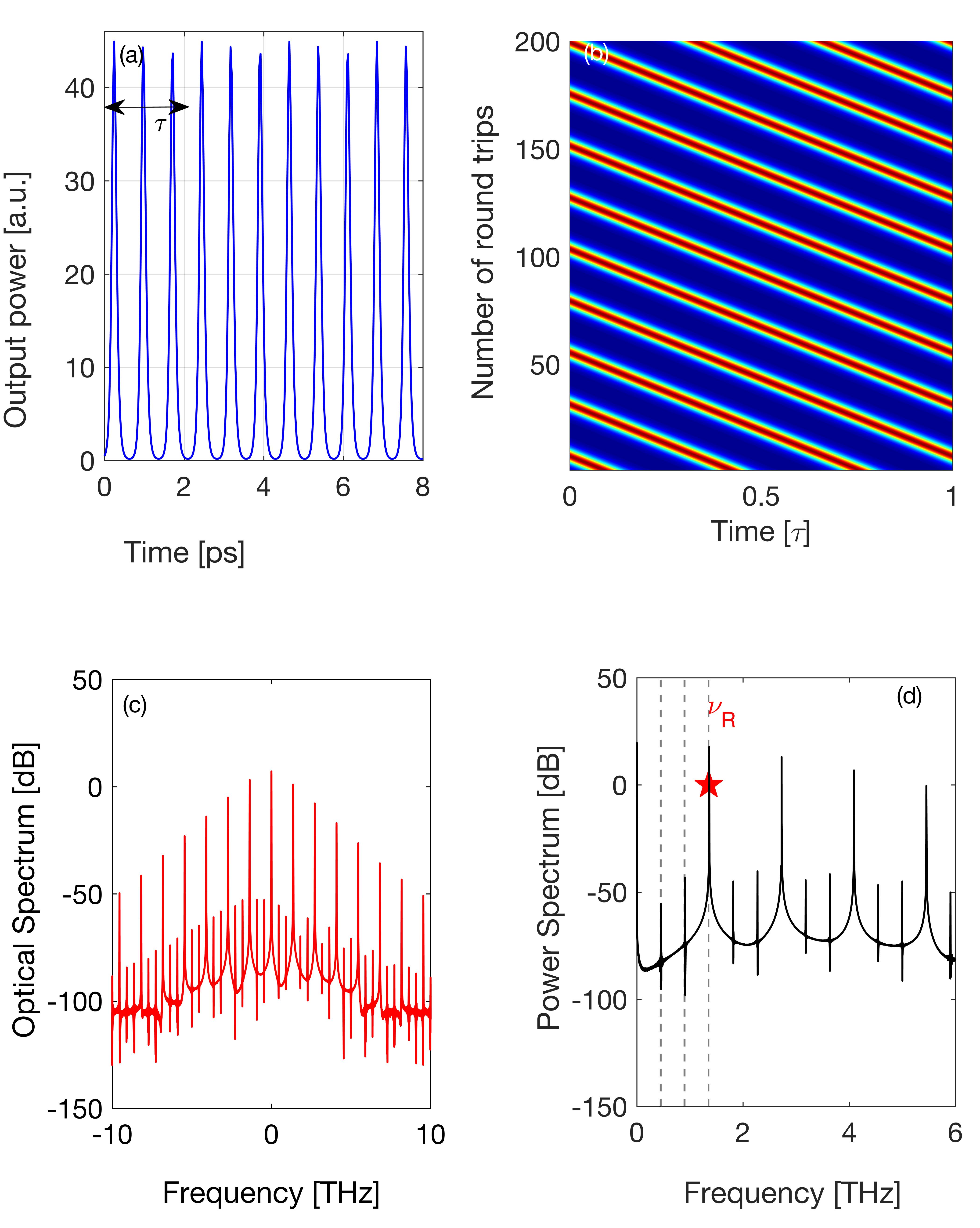}
 \end{center}
\caption{Temporal evolution of the output power (a,b), optical spectrum (c) and RF spectrum (d) obtained for a value of bias current of \SI{75} {\milli \ampere}. 
In panel (b) we report a space-time representation of the pulse dynamics for \SI{75} {\milli \ampere}. A long time trace is divided in intervals corresponding to the cold cavity round trip time $\tau=L\eta/c$. These segments are then stacked on top of each other so that the horizontal axis is equivalent to space inside the cavity while the vertical dimension describes the evolution in units of round trips. The lowest frequency dashed line in panel (d) corresponds to the ring resonator FSR (equal to $\simeq$ \SI{440} {\giga \hertz}), while the other two dashed lines are integer multiple of it. The other parameters are those used in Fig. \ref{fig0}.}
\label{num2}
\end{figure}
 \begin{figure}[h!tb]
 \begin{center}
\includegraphics[width=0.8\textwidth]{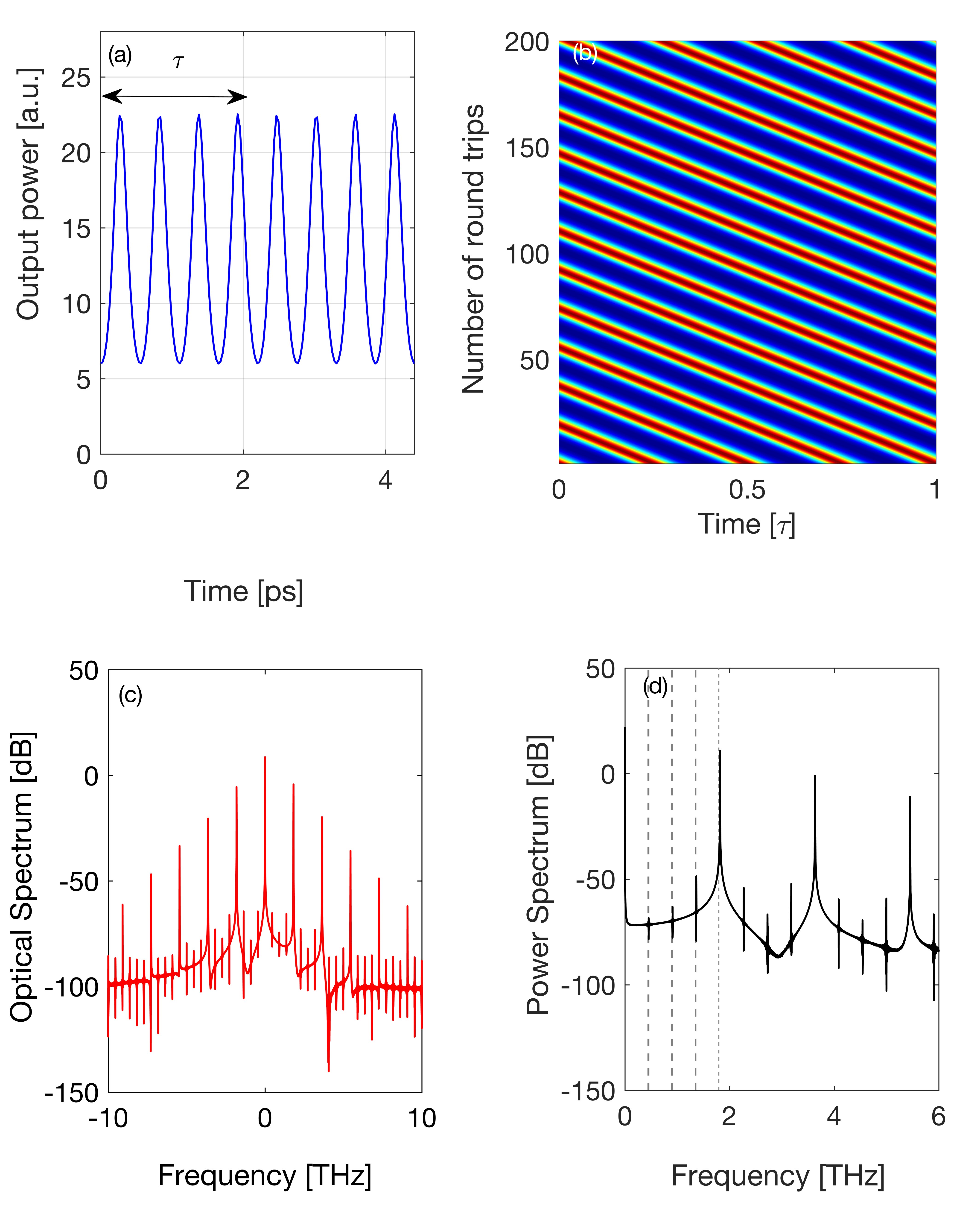}
 \end{center}
\caption{Temporal evolution of the output power (a,b), optical spectrum (c) and RF spectrum (d) obtained for a value of bias current of \SI{95} {\milli \ampere}. Dashed lines in panel (d) denote the first cold cavity modes. The other parameters are those used in Fig. \ref{num2}.}
\label{num2b}
\end{figure}

 We integrate the TDTW model equations using a finite difference algorithm as described in \cite{Gioannini, Bardella2017}.\\
 
\begin{figure}[h!tb]
 \begin{center}
\subfigure[]{\includegraphics[width=0.49\textwidth]{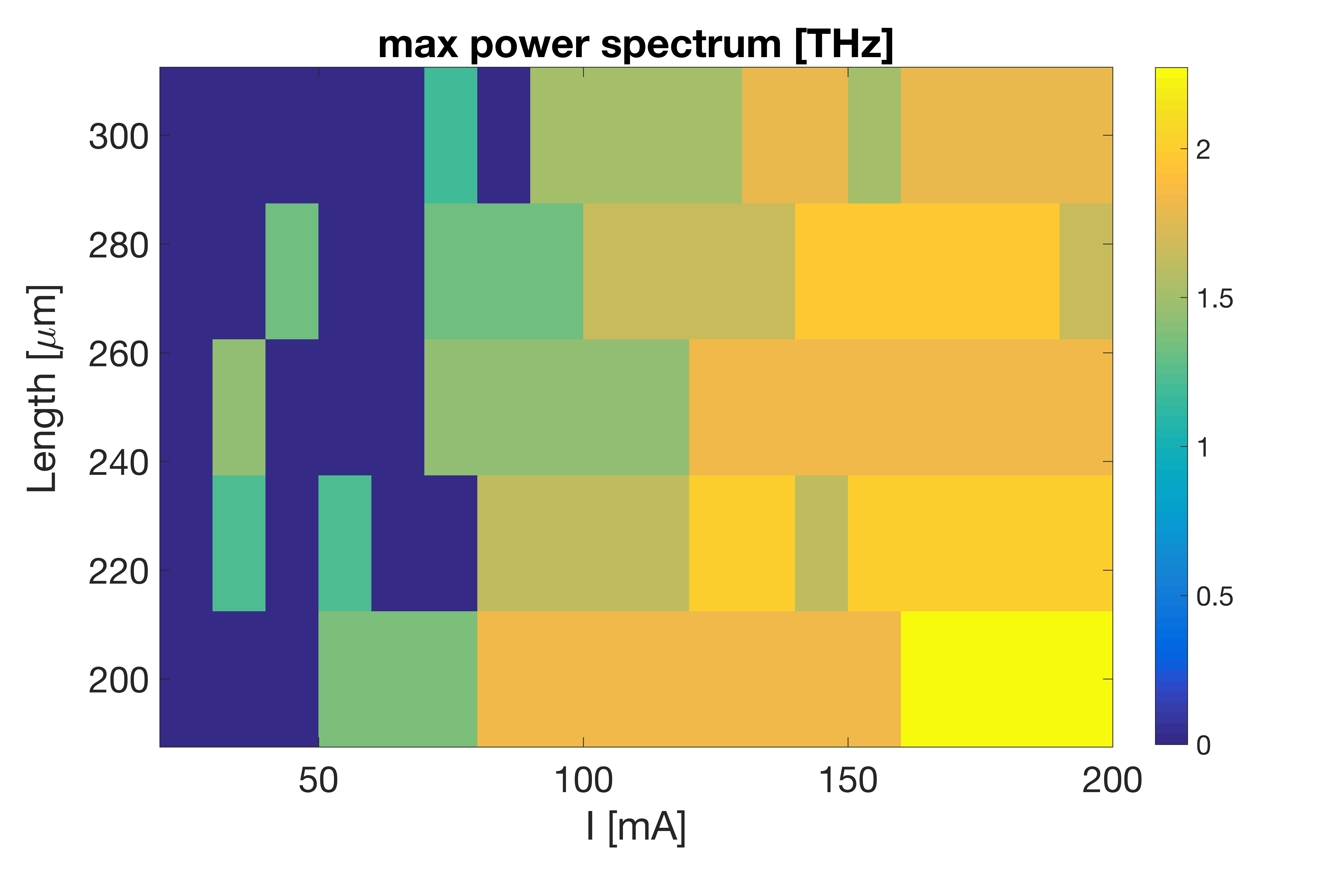}}
\subfigure[]{\includegraphics[width=0.46\textwidth]{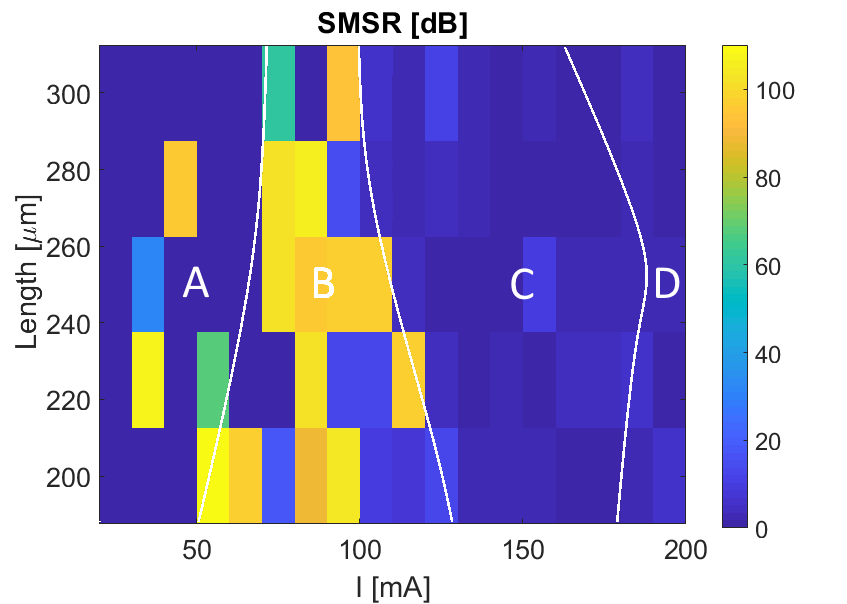}}
 \end{center}
\caption{Colour map representation of the frequency corresponding to the absolute maximum in the RF spectrum (a) ($\nu$ $\ne$ $0$), always close to $\nu_{R}$, and the ratio between its amplitude and that of the highest side mode as a function of the cavity length $L$ and the bias current $I$. The other parameters are those used in Fig. \ref{num2}.}
\label{num4}
\end{figure}

For the parameters in Table \ref{Tableparam} we obtained the bifurcation diagram in Fig. \ref{fig1} where the maxima and minima in the intensity time traces are reported versus the bias current as control parameter. Red lines correspond to the TW solutions calculated using Eqs. (\ref{Pfastaz})-(\ref{pop1fstaz}). As predicted by the linear stability analysis, for $I\ge 55mA$ the TW solution becomes unstable. In particular the multimode competition gives rise to regular intensity oscillations. The number of excited modes and the pulse contrast both increase with bias current. In Fig. \ref{num2} we report for e.g. the temporal evolution of the output power, the optical spectrum and RF spectrum for $I=75 mA$. In this case the first unstable mode has a distance of approximately $3$ times the cavity FSR ($\simeq$ \SI{440} {\giga \hertz}) with respect to the TW emission frequency as shown in Fig. \ref{num2}.d and, in perfect agreement with the results of the LSA in Fig. \ref{fig0}, it corresponds to the first lasing mode with a positive parametric gain. The side mode suppression ratio (SMSR) defined here as the ratio between the maximum RF peak power to that of the highest adjacent longitudinal modes is around $60$ $dB$.
Moreover, because of the amplitude character of RNGH instability \cite{Lugiato}, the phases of the first excited longitudinal modes are locked with equal phase difference between adjacent modes. Once the first side modes are activated, a cascaded Four Wave Mixing (FWM) mechanism comes into play in fixing the frequency and the phase of the parametrically generated modes, thus yielding to the emission of ultra-short pulses at THz emission rate (see Fig. \ref{num2}.a). In the useful spatio-temporal representation in Fig. \ref{num2}.b three pulses are associated with a single cold cavity round trip time $\tau=L\eta/c$. As expected by the results of the LSA and shown for e.g. in Fig. \ref{num2b}, the increase of the bias current allows us to partially tune the pulse repetition rate by changing the parametric gain peak position, or equivalently, the Rabi frequency of the system. 
In order to demonstrate the robustness of the SP phenomenon against ring length and current variation we run a set of systematic simulations. Our results might be conveniently summarised in Fig. \ref{num4} as a function of the cavity length $L$ and the bias current $I$. In Fig. \ref{num4}.a we map the frequency of the RF peak (that turns to be always close to the Rabi frequency $\nu_{R}$). In Fig. \ref{num4}.b, in order to evaluate the spectral purity of the pulsed THz signal, we report  the ratio between the power of the RF peak and the RF power of the competing adjacent RF lines corresponding to the ring modes not triggered by the RNGH instability.  We define this ratio in the RF spectrum as SMSR (see Fig. 4d). It is possible to identify at least four different dynamical behaviour. In the great part of region A we have no pulsing with only one lasing line and CW power (TW stable). In the region denoted by the letter B the QD ring laser shows SP at a frequency in the THz range close to Rabi resonance, in the region denoted by the letter C phase-locking still induces regular oscillations although the emergence of side modes introduces pulse over-modulation, and finally in the region denoted by the letter D the multimode dynamics leads to irregular oscillations. 

 
\begin{figure}[h!tb]
 \begin{center}
\includegraphics[width=0.8\textwidth]{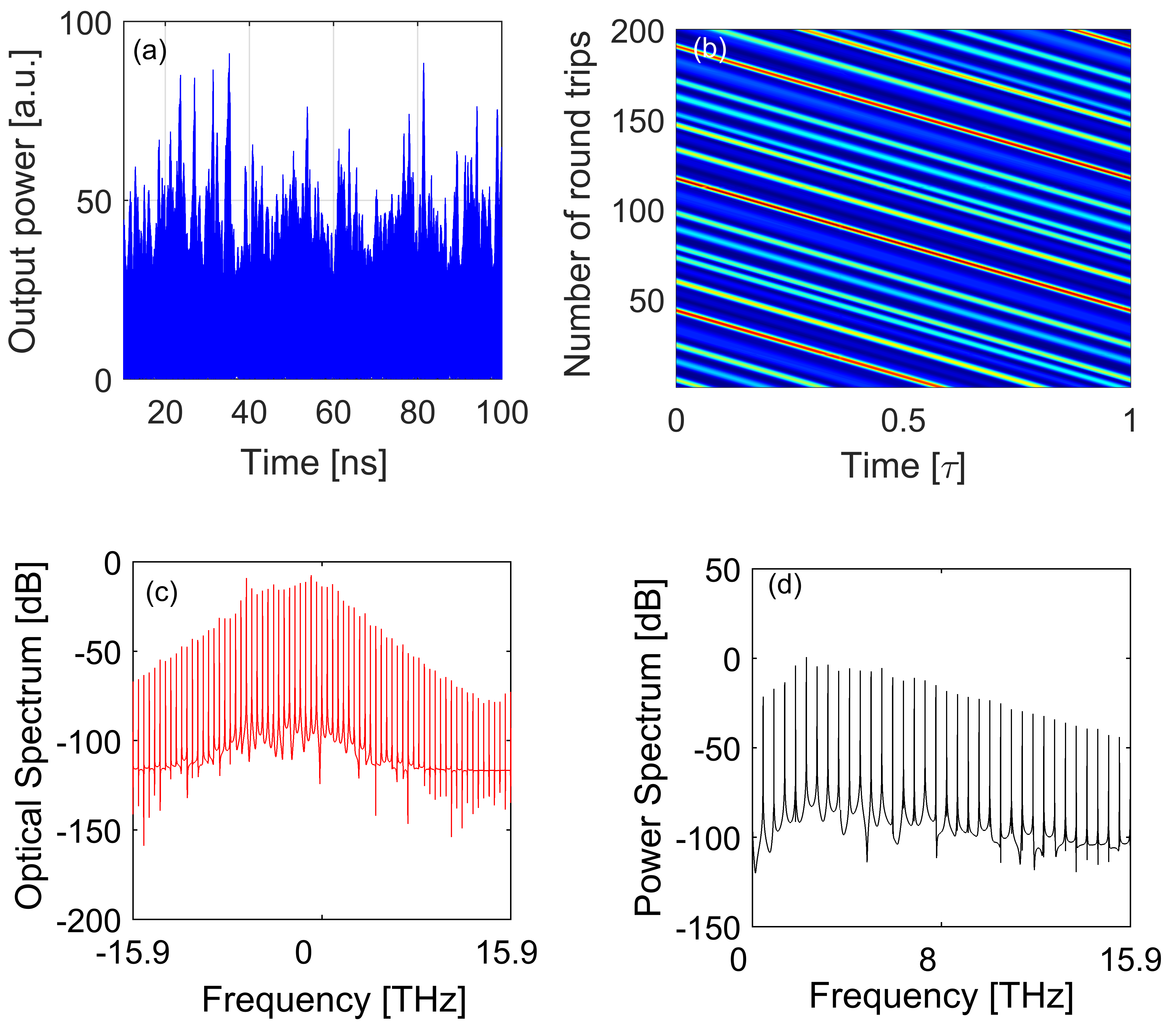}
 \end{center}
\caption{Temporal evolution of the output power (a,b), optical spectrum (c) and RF spectrum (d) obtained for a value of bias current of \SI{75} {\milli \ampere}. We consider $11$ QDs populations whose central emission frequencies are again separated by a \SI{1} {\tera \hertz}. The other parameters are those used in Fig. \ref{num2}.}
\label{num3}
\end{figure}

\begin{figure}[h!tb]
 \begin{center}
\includegraphics[width=0.8\textwidth]{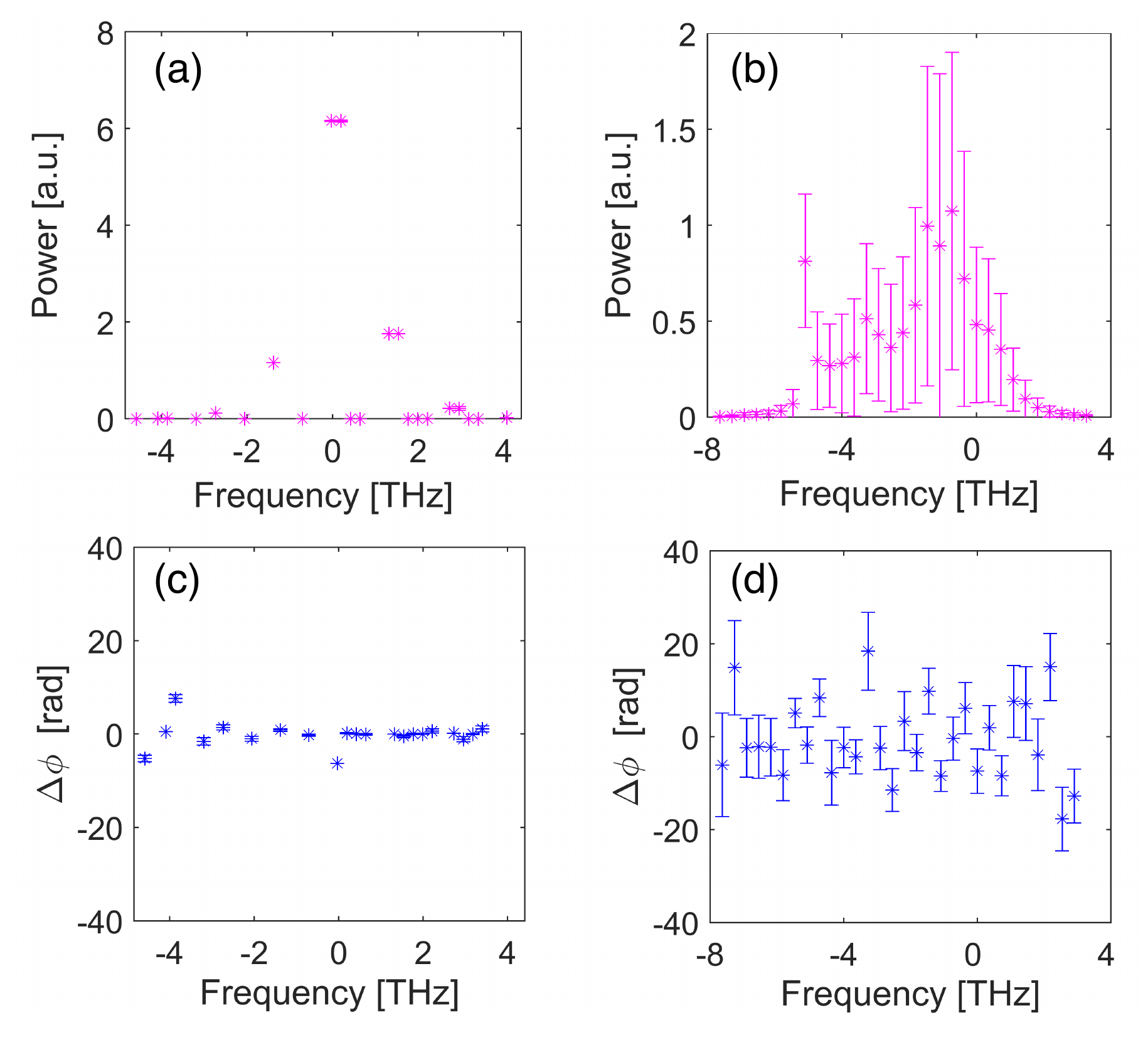}
 \end{center}
\caption{Average values of the modal amplitude and of the standard deviation of the modal phase. The errors bars denote the standard deviation of their temporal fluctuations in the case of $3$ (a), (c) and $11$ populations (b), (d). The other parameters are those used in Fig. \ref{num2}.}
\label{num2c}
\end{figure}

Our simulations also show that an increase of the inhomogeneous broadening in the model reduces the intervals of bias current where SP is found.
We might expect in fact that excited longitudinal modes with different frequencies interact with different populations thus reducing the degree of coherence in the system. In Fig. \ref{num3} we plot for e.g. the results obtained by considering $11$ populations whose central emission frequencies are separated by $1THz$ and that correspond to a FWHM of the effective inhomogeneous broaden gain linewidth of $\simeq$ \SI{10} {\tera \hertz} ($\simeq$ \SI{41} {\milli \electronvolt}) (see Fig.\ref{fig:3}), while keeping the other parameters as those in Fig. \ref{num2}. In this case we observe an irregular temporal evolution of the field intensity that corresponds to a much higher differential phase and amplitude noise (see Fig. \ref{num2c}). At the same time multimode emission experience a reduced threshold and a larger bandwidth because the material gain for non resonant (with the gain peak) modes gets higher \cite{Lugiato}.

 \subsection{Bidirectional ring}
 
\begin{figure}[h!tb]
 \begin{center}
\includegraphics[width=0.8\textwidth]{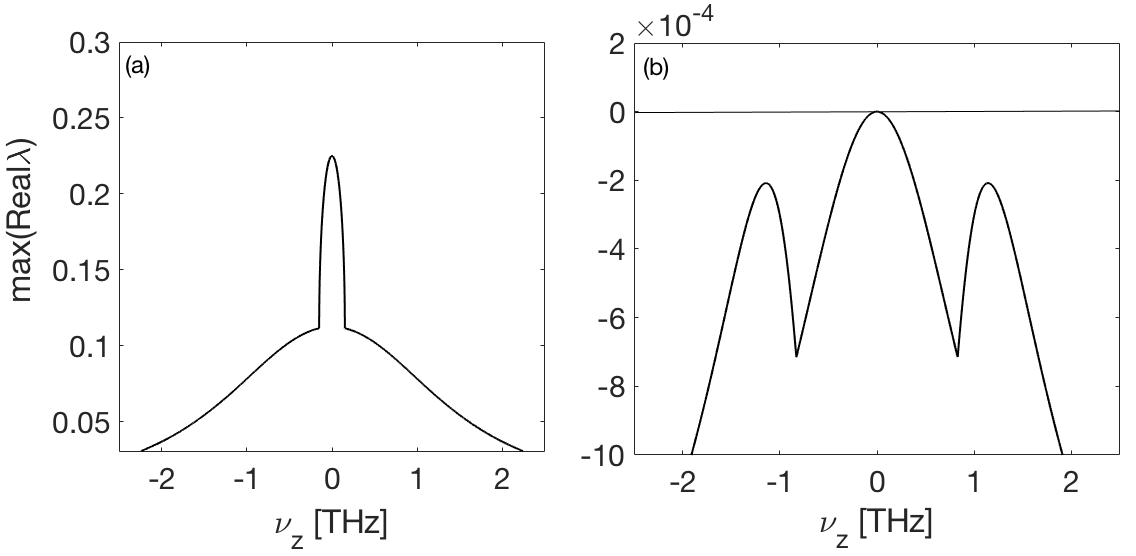}
 \end{center}
\caption{Results of the LSA of the TW solutions for $I=60 mA$ for a bidirectional ring configuration (a) and an unidirectional one (b). Plot of the parametric gain for each value of the frequency $\nu_{z}=\omega_{z}/2 \pi=k_{z}\,v_{g}/2 \pi$ treated as continuous variable. The other parameters are those used in Fig. \ref{num2}.}
\label{num2e}
\end{figure}

We finally observe that in the bidirectional configuration, the standing wave pattern due to the interference between forward and backward fields generates a grating in the carrier density that cannot be washed out by diffusion. Equations for the first Fourier components of the spatial grating are added following the procedure described in \cite{Bardella2017}. Spatial hole burning takes place, letting the TW instability threshold decrease from several times the lasing threshold down to a few percents above the lasing threshold. This emerges for e.g. from inspection of Fig. \ref{num2e} where, focusing on the simple case of a single QD population (homogeneous gain broadening), we report the linear stability analysis of the TW solutions for a bidirectional configuration (panel a) and a unidirectional one (panel b). The latter has been carried on via a calculations analogous to that described in  \cite{Boiko, Gordon} in the case of a QCLs.
In the unidirectional configuration only the mode at \SI{0} {\tera \hertz} has positive parametric gain (see Fig. \ref{num2e}.a); all the others are suppressed and only by increasing current the two relative maxima at $\simeq$ $\pm$ \SI{1.2} {\tera \hertz} will experience positive parametric gain and they will allow the lasing of the modes closer to these two maxima.
Instead, in the bidirectional ring configuration all the cavity modes in the frequency range of few \SI{}{\tera \hertz} experience a positive parametric gain (see Fig. \ref{num2e}.a) which is turned-on by the spatial hole burning effect. The TW resonant with the gain peak is unstable very close to the lasing threshold and by increasing the bias current we generally observe an alternation between regimes of irregular oscillations and a regular dynamical behavior as recently reported in \cite{Bardella2017}. Coherent dynamics leading to self-generation of OFCs with lasing lines spaced of the ring FSR is found for sizeable intervals of the bias current. It does not correspond to the emission of optical pulses (since the phase difference between adjacent modes is not equal), although is normally associated to the emission of a broader and flatter optical spectrum. \\

\section{Conclusions}

We studied self-mode-locking and in particular self-pulsing in single section ring QDs lasers. In unidirectional emission regime ultra-short pulses at THz repetition rate are triggered by RNGH multi-wavelengths instability of the TW solutions that consists in the amplification of the Rabi frequency of the system. The latter has been calculated in the very well verified hypothesis that radiation coherently interacts with QDs material as with an ensemble of artificial two level atoms.
In bidirectional cavities, spatial hole burning makes the TW instability threshold occurring for much lower bias current, but only self-generation of OFCs is reported. Our results let envisage very timely applications such as the high-data rate optical information encoding and transmission or the generation of THz or sub-THz signals via combination of photonics and electronics. 

\section{Appendix A}

We study the stability of the TW emission respect to spatio-temporal perturbations looking for solutions of Eqs. (\ref{fieldfastc})-(\ref{pop2fastb}) in the form:
$$E=(\overline{E}+\delta E)e^{j(\delta \omega/\Gamma \, t- \delta k \, L z)} \quad p_{i}=(\overline{p_{i}}+\delta p_{i})e^{j(\delta \omega/\Gamma \, t- \delta k \, L z)}$$$$\rho_{WL}(z,t)=\overline{\rho_{WL}}+\delta \rho_{WL} \quad \rho_{i}(z,t)=\overline{\rho_{i}}+ \delta \rho_{i}$$
This gives the following set of linear equations for the perturbations: 
\begin{eqnarray}
\frac{\partial \delta E}{\partial t} + \gamma_{p}\frac{\partial \delta E}{\partial z}&=&\gamma_{p}\left(-\frac{\alpha_{wg}L}{2}\delta E-C\sum_{i=-N}^{N} \bar{G}_{i} \delta p_{i} \right)\label{deltafieldfastc}\\
\frac{\partial \delta p_{i}(z,t)}{\partial t}&=&(j\delta_{i}/ \Gamma-1-j \delta \omega /\Gamma) \delta p_{i}-D(2\delta \rho_{i})E-D(2\rho_{i}-1) \delta E \label{deltaPfast1b}\\
\frac{\partial \delta \rho_{i}(z,t)}{\partial t}&=& -\delta \rho_{i}\gamma_{e}(1-\rho_{WL})+ \rho_{i}  \delta \rho_{WL}\gamma_{e} -F \delta \rho_{i} \rho_{WL} \gamma_{C} \nonumber \\
&+&  F  (1-\rho_{i})\delta \rho_{WL} \gamma_{C} - 2 \rho_{i}  \delta  \rho_{i}+ H \, Re\left(\delta E^{*}p_{i}+ E^{*}\delta p_{i}\right)         \label{deltapop1fastb}\\
\frac{\partial \delta \rho_{WL}(z,t)}{\partial t}&=&-\delta \rho_{WL} \gamma_{nr}^{WL}+\sum_{i=-N}^{N}\left[-\bar{G}_{i} \delta \rho_{WL}\gamma_{C}(1-\rho_{i}) \right.\nonumber\\ 
&+&  \left. \bar{G}_{i}  \rho_{WL}\gamma_{C} \delta \rho_{i}+\frac{\bar{G}_{i}}{F} \delta \rho_{i}\gamma_{e}(1-\rho_{WL}) - \frac{\bar{G}_{i}}{F}\rho_{i}\gamma_{e}\delta \rho_{WL} \right] \label{deltawettingpop2fastb}
\end{eqnarray}
Projecting on the spatial Fourier basis the perturbations we derive for each perturbation wave vector $k_{n}$ a set of ODE for the temporal evolution of the corresponding Fourier component. The maximum of the real parts of the eigenvalues $\lambda$ of the associated Jacobian matrix (Lyapunov exponents), representing the parametric gain of the considered mode, thus give a direct information about the TW stability.

\end{document}